\documentclass[fleqn,10pt]{wlscirep}
\usepackage{hyperref}

\title{Reply to "The equivalence of the Power-Zineau-Woolley picture and the Poincar\'e gauge from the very first principles" by G. K\'onya \textit{et al.}\cite{2018arXiv180105590K}}

\author[1,*]{Emmanuel Rousseau}
\author[1]{Didier Felbacq}

\affil[1]{Universit\'e de Montpellier , Laboratoire Charles Coulomb UMR 5221, F-34095, Montpellier, France}

\affil[*]{emmanuel.rousseau@umontpellier.fr}

\keywords{Keyword1, Keyword2, Keyword3}

\begin{abstract}
This note is a reply to the paper \cite{2018arXiv180105590K} arXiv:1801.0559: "The equivalence of the Power-Zineau-Woolley picture and the Poincar\'e gauge from the very first principles" by G. K\'onya \textit{et al}.

--------------------------------------

In a recent paper\cite{rousseau}, we have shown that the Power-Zienau-Woolley Hamiltonian does not derived from the minimal-coupling hamiltonian with the help of a gauge transformation. This result has been challenged by G. K\'onya \textit{al.} in a comment \cite{2018arXiv180105590K} where the authors claim the equivalence between the Power-Zienau-Woolley hamiltonian\cite{Power1959,Woolley1979, Babiker1983,PhysRevA.28.2649,Cohen} and the minimal-coupling hamiltonian in the Poincar\'e gauge. 
They claim that we have made one error and one wrong emphasis in our paper:

\textit{\textbf{The error as summarized by G. K\'onya \textit{al.} would be}: "The canonical field momentum is not gauge invariant. Equivalent transformations of the Lagrangian do change the momentum. In field theories, gauge transformations are special cases of such transformations. The electric field $\vec{E}$ is gauge invariant, but its capacity of being the canonical momentum is not."}

\textit{\textbf{The wrong emphasis as summarized by G. K\'onya \textit{al.} would be}: "The use of the canonical coordinate/momentum pair $\vec{A}_p$ and $\vec{E}$ in Poincar\'e gauge is presented as mandatory in Rousseau and Felbacq paper, whereas as there is a certain freedom of choice in selecting this pair. Also in Poincar\'e gauge it is possible to use $\vec{A}_c$ as canonical coordinate, in which case the conjugate momentum will be $\vec{D}$. This is the most convenient choice in terms of the set of nontrivial Dirac brackets. Cf. Table 1 in G. K\'onya \textit{al.} paper\cite{2018arXiv180105590K} for possible choices."}

--------------------------------------

We do not share these conclusions and show in this reply that these statements are incorrect. Specifically, we show that under a gauge transformation, the canonical momentum $\pi(\vec{x},t)$ conjugated to the vector potential $\vec{A}(\vec{x},t)$ is given by $\pi(\vec{x},t)=-\varepsilon_0 \vec{E}(\vec{x},t)$. 

This happens because the Lagrangian does not contains terms proportional to $\partial_t \phi(\vec{x},t)$ where $\phi(\vec{x},t)$ is the scalar potential.

Moreover our choice of canonical variables was challenged. Actually, our set of independent variables is exactly the same as in G. K\'onya \textit{al.} \cite{2018arXiv180105590K} except that we do not write explicitly the dependent variables in term of the independent ones. This is one great advantage of the Dirac procedure for constrained hamiltonian\cite[p.347]{WeinbergField}.
\end{abstract}

\begin{document}

\flushbottom
\maketitle
%
%
\thispagestyle{empty}

%
%
%


\section{The canonical momentum conjugated to $\vec{A}_p/\vec{A}_p^\bot$ is $-\varepsilon_0 \vec{E}$}

In order to recover the Power-Zienau-Woolley Hamiltonian, it is mandatory to find that the canonical momentum $\vec{\pi}$ conjugated to the vector potential is the displacement vector $\vec{D}$, \textit{i.e.} $\vec{\pi} = -\vec{D}$. We show in the following that cannot be correct. We do the calculations in two ways. The first one follows the comment by G. K\'onya \textit{al.}\cite{2018arXiv180105590K} in writing the explicit dependence between the dynamical variables (see subsection \ref{expl} ). The second way (subsection \ref{impl} ) follows the Dirac procedure for constraint hamiltonians. This method highlights unambiguously the contribution of the constraint $\partial_t \pi_\phi=0$. In both ways, we show that the canonical momentum$\vec{\pi}$ equal the electric field $\vec{\pi} = -\varepsilon_0 \vec{E}$ because of the constraints $\pi_\phi=0$ and $\partial_t \pi_\phi=0$ coming from the fact that the Lagrangian is free from $\partial_t \phi(\vec{x},t)$-term.

For completeness let us first recall our main assumptions and the starting points share by G. K\'onya \textit{al.}\cite{2018arXiv180105590K} and our work\cite{rousseau}. We consider one single electron with position $\vec{r}$ and electric charge $q$ evolving in a binding potential $V(\vec{r})$. The electron interacts with the electromagnetic field. The electromagnetic-field dynamical-variables are the vector potential $\vec{A}(\vec{x},t)$ and the scalar potential $\phi(\vec{x},t)$. The gauge is fixed to the Poincar\'e gauge defined by $\vec{x}.\vec{A}_p(\vec{x},t)=0$ for all points $\vec{x}$ in space.

In the Poincar\'e gauge the Lagrangian reads as

\begin{eqnarray}
L_p(\vec{r}, \vec{A}_p,\phi_p) &=&  \frac{1}{2}m\dot{\vec{r}}^{~2} - V(\vec{r}) \label{eq:Lpoin1}  \\
&+& \int d\vec{x} \frac{1}{2}\varepsilon_0[(\partial_t \vec{A}_p(\vec{x},t)+\nabla \phi_p(\vec{x},t))^2 - \frac{1}{2\mu_0}(\vec{\nabla} \times \vec{A}_p(\vec{x},t))^2]  \\
&+& q\dot{\vec{r}}~.~\vec{A}_p(\vec{r},t)-q\phi_p(\vec{r})
\label{eq:Lpoin3}
\end{eqnarray}

This is the eq.(38) in ref.\cite{2018arXiv180105590K} except that these authors have expressed the potentials with their values in the Poincar\'e gauge:

\begin{eqnarray}
\vec{A}_p(\vec{r},t) &=& - \vec{r} \times \int_0^1 udu \vec{B}(u\vec{r}) =   - \vec{r} \times \int d\vec{x} \int_0^1 udu \vec{B}(\vec{x}) \delta(\vec{x} - u\vec{r}) \label{eq:Ap} \\
\phi(\vec{r},t)_p &=& - \vec{r} . \int_0^1 du \vec{E}(u\vec{r}) = - \vec{r} . \int d\vec{x} \int_0^1 du \vec{E}(\vec{x}) \delta(\vec{x} - u\vec{r})\label{eq:phip}
\end{eqnarray}

To understand the discrepancies between G.K\'onya \textit{et al.} results \cite{2018arXiv180105590K}  and our results\cite{rousseau} one should have in mind the similarities and the differences between our two works. Our approach\cite{rousseau} is based on the Dirac procedure for constrained hamiltonians\cite{Dirac,henneaux} whereas the approach by  K\'onya \textit{et al.} \cite{2018arXiv180105590K} writes explicitly the constraints between the dynamical variables. Similarities and discrepancies are summarized in the table (\ref{table:comp}). This table shows that the two approaches share similar inputs except one constraint that is missing in K\'onya \textit{et al.} \cite{2018arXiv180105590K}. This constraint imposes that the canonical momentum $\pi_\phi$ associated to the scalar potential $\phi$ has to remain null at any time. \emph{ This is the origin of the discrepancy between our results}.

\begin{table}
\begin{center}
\begin{tabular}{|c|c|c|} 
   \hline
   Physical quantities & Rousseau and Felbacq\cite{rousseau} & K\'onya \textit{et al.} \cite{2018arXiv180105590K} \\
   \hline
   Lagrangian & \multicolumn{2}{c|}{The same given by eq.(\ref{eq:Lpoin1}-\ref{eq:Lpoin3})} \\
    \hline
    Constraint; $\chi_1 = \pi_\phi=0$ & Yes & Yes \\
    \hline
     Constraint; $\chi_2 = \partial_t \pi_\phi $ & Yes & \textcolor{red}{No} \\
     $\chi_2 =  -[q \delta(\vec{x}-\vec{r}) + \vec{\nabla}.\vec{\pi}(\vec{x},t)]=0$ &  &  \\
     \hline
     Constraint;  & Yes & Yes, through vector potential equation;  \\
     $\chi_3 =\vec{x}.\vec{A}(\vec{x},t)=0$ &  & eq.(\ref{eq:Ap}) \\
     \hline
     Constraint; $\chi_4 = \vec{x}.\partial_t \vec{A}(\vec{x},t)  $ & Yes & Yes, through scalar potential  equation\\
      $\chi_4 = \vec{x}.[\frac{\vec{\pi}}{\varepsilon_0} - \vec{\nabla} \phi(\vec{x},t)]=0$ &  & eq.(\ref{eq:phip}) \\
     \hline
     & Dirac procedure  & Explicit writing of the constraints  \\
     methods for constrained hamiltonian & and  & at the Lagrangian level  \\
      & Dirac Brackets &   \\
      \hline
      \textcolor{red}{Discrepancy}: & $-\varepsilon_0 \vec{E}(\vec{x},t)$ & $-\vec{D}(\vec{x},t)$  \\
      Canonical momentum $\vec{\pi}$ &  &  \\
      &	 electric field & displacement field \\
    \hline
\end{tabular}
 \caption{Similarities and differences in Rousseau and Felbacq results\cite{rousseau} and in K\'onya \textit{et al.} \cite{2018arXiv180105590K}. The main difference lyies in the method used to get the hamiltonian. Rousseau and Felbacq\cite{rousseau} used the Dirac theory for constrained hamiltonien whereas K\'onya \textit{et al.} \cite{2018arXiv180105590K} wrote explicitly the constraints at the level of the Lagrangian.}
  \label{table:comp}
 \end{center}
\end{table}

Although, both papers share lots of similarities, the final results are different: this comes from a different canonical momentum $\vec{\pi}$. In order to understand the origin of the discrepancy let us emphasize the main advantages of Dirac theory. In this theory, although the dynamical variables may be dependent upon each other, they are considered as being independent variables, while their interdependences are taken into account through constraints used to compute the Dirac Brackets. The Dirac brackets add corrective terms to the Poisson brackets, these corrective terms arising from the constraints. Among several advantages, the Dirac theory is algorithmic in the sense that the procedure is based on theorems\cite{henneaux}. If the procedure is done in a correct way, a unique solution exists for the transformation from the Lagrangian formalism to the Hamiltonian formalism\cite{Dirac, henneaux}. On another hand, from a more a concrete point of view, an important consequence of the theory is that the dynamical variables are considered as being independent from each other. As a consequence, one has not "to use explicit expressions for the dependent variables in terms of the independent ones"\cite[p.347]{WeinbergField}. In such a way, taking the functional derivative of any quantity can be done safely without having to take into account for the constraints between the variables.

When computing the canonical momentum conjugated to the vector potential, using Dirac theory as we have done, we can compute the functional derivative safely since all dynamical variables are assumed to be independent from each other. In other words, in Dirac theory, the functional derivative has to be understood as $ \frac{\delta L_p}{\delta \partial_t A_p} = \frac{\delta L_p}{\delta \partial_t A_p}\big |_{\phi_p=cst}$. The previous notation means that the scalar potential $\phi_p$ is considered as a constant with respect to the functional derivative. In such a case, we found\cite{rousseau}: 

\begin{eqnarray}
\vec{\pi} =  \frac{\delta L_p}{\delta \partial_t A_p}\bigg |_{\phi_p=cst} = -\varepsilon_0 \vec{E}
\label{eq:piR}
\end{eqnarray}

In G. K\'onya \textit{et al.}'s approach the difficulty is to perform variations in phase-space only on the manifold allowed by the constraints. Following their approach, the transverse part of the vector potential $\vec{A}^\bot_p$ is the dynamical variable. The longitudinal part of the vector potential $\vec{A}^\parallel_p$ and the scalar potential $\phi_p$ are considered as functions of the transverse part. Concerning the canonical momentum, they found: 

\begin{eqnarray}
\vec{\pi}' = -\varepsilon_0 \vec{E} -q\vec{r}\int_0^1 du \delta(\vec{x}-u\vec{r}) 
\label{eq:piK}
\end{eqnarray}

As in our result, the first term originates from variations of the term $\frac{\varepsilon_0}{2}\vec{E}^2$ in the Lagrangian. The second term arises from variation of the term $- \vec{r} . \int_0^1 du \vec{E}(u\vec{r})$ that is nothing else but the scalar potential in the Poincar\'e gauge [see eq.(\ref{eq:phip})]. Indeed $q \frac{\delta}{\delta \partial_t A_p^\bot} [- \vec{r} . \int_0^1 du \vec{E}(u\vec{r})] = q\int_0^1 du \delta(\vec{x}-u\vec{r}) $. To summarize, the canonical momentum $\vec{\pi}'$ computed by K\'onya \textit{et al.} \cite{2018arXiv180105590K} includes also variations along the scalar potential variable. 

\begin{eqnarray}
\vec{\pi}' = \frac{\delta L_p}{\delta \partial_t A_p^\bot}\Bigg |_{\phi_p=cst} + \frac{\delta L_p}{\delta \phi_p} \Bigg |_{A_p^\bot=cst}\frac{\delta \phi_p[A_p^\bot]}{\delta \partial_t A_p^\bot} 
\nonumber
\end{eqnarray}

In the next subsection, we compute the canonical momentum $\vec{\pi}'$ one the manifold allowed by the constraints.

\subsection{Calculation of the canonical momentum conjugated to $\vec{A}$ with an explicit writing of the constraints \label{expl}}

To do so, we consider the Lagrangian given by the set of equations (\ref{eq:Lpoin1}-\ref{eq:Lpoin3}). The scalar potential $\phi_p(\vec{x},t)$ is expressed with the help of the equation (\ref{eq:phip}). The longitudinal part of the vector potential $\vec{A^\parallel_p}$ is given in the Poincar\'e gauge by the gauge-generating function. It reads $\vec{A^\parallel_p} = - \vec{\nabla} \int_0^1 du~ \vec{x}.\vec{A^\bot_p}(u\vec{x},t)$. We are actually assuming a transition from the Coulomb gauge to the Poincar\'e gauge. 
As a consequence, we consider the transverse part of the vector potential $\vec{A^\bot_p}(\vec{x},t)$ as the only dynamical variable for the electromagnetic field. Here we are mimicking K\'onya \textit{et al.} \cite{2018arXiv180105590K}. We are assuming an explicit dependence of the longitudinal part of the vector potential $\vec{A}^\parallel[\vec{A}^\bot]$ and of the scalar potential $\phi[\vec{A}^\bot]$ with the transverse part of the vector potential. 

\begin{eqnarray}
\vec{\pi}' = \frac{\delta L_p}{\delta \partial_t A_p^\bot}\Bigg |_{\phi_p=cst} + \frac{\delta L_p}{\delta \phi_p} \Bigg |_{A_p^\bot=cst}\frac{\delta \phi_p[A_p^\bot]}{\delta \partial_t A_p^\bot} 
\nonumber
\end{eqnarray}

The functional derivative occurs along a path for which the action is extremal. Then the Euler-Lagrange equation holds 
$$\frac{\delta L_p}{\delta \phi}\Bigg |_{A_p^\bot=cst} =\partial_t \frac{\delta L_p}{\partial_t \delta \phi}\Bigg |_{A_p^\bot=cst}$$

 One finds $\frac{\delta L_p}{\delta \phi}\big |_{A_p^\bot=cst} = -\varepsilon_0\vec{\nabla}.(\partial_t \vec{A}_p + \vec{\nabla} \phi_p) - q \delta(\vec{x}-\vec{r}) = 0$. This is the Maxwell-Gauss equation. But, most importantly for our purpose, this equation is a consequence of the first constraint $\chi_1= \pi_\phi=0$. Indeed since this constraint has to hold at any time $\partial_t \pi_\phi = \partial_t \frac{\delta L_p}{\partial_t \delta \phi}|_{A_p^\bot=cst} = 0$. If one wishes that the system remains on the surface defined  by $\pi_\phi = 0$ at any time, one must have $\frac{\delta L_p}{\delta \phi}|_{A_p^\bot=cst} =0$. The canonical momentum conjugated to $A_p^\bot$ is then:

\begin{eqnarray}
\vec{\pi}' = \frac{\delta L_p}{\delta \partial_t A_p^\bot}\Bigg |_{\phi_p=cst} = \vec{\pi} 
\nonumber
\end{eqnarray}

It reduces to the same equation as ours [eq:(\ref{eq:piR})]. As a consequence of the constraint $\partial_t \pi_\phi = 0$, the functional derivative must be evaluated as if the scalar potential is a constant. The result $\vec{\pi}' = \vec{\pi} = -\varepsilon_0 \vec{E}$ is recovered as in our paper\cite{rousseau} but following K\'onya \textit{et al.} \cite{2018arXiv180105590K} method.
Because the scalar potential and the transverse part of the vector potential are coupled through the equation (\ref{eq:phip}) a small variation $  \delta \partial_t \vec{A_p^\bot}$ induces a variation of the time derivative of the scalar potential $ \delta \partial_t \phi_p$. In the hamiltonian formalism this last variation $\partial_t \delta \phi_p$ implies a variation of the canonical momentum $\delta \pi_\phi$. But $\pi_\phi$  and its variations are constrained. So $\delta \pi_\phi$ must be null, \textit{i.e.} $\delta \pi_\phi= 0$. This is the error done by K\'onya \textit{et al.} \cite{2018arXiv180105590K}. They did not realize that the term $-q\int_0^1 du \delta(\vec{x}-u\vec{r})$ arises from variations of the scalar potential. They have differentiated the term arising from the scalar potential in the Pioncar\'e gauge  as if it were an independent term, which is not.

To be exhaustif, as noted by G. K\'onya \textit{al.}\cite{2018arXiv180105590K}, there is an ambiguity in the definition of the canonical momentum $\vec{\pi}(\vec{x},t)$ [see also ref.\cite[p.348]{WeinbergField} for more details]. As a matter of fact, changing $ \partial_t \vec{A_p^\bot}(\vec{x},t)$  by the amount $\delta \partial_t \vec{A_p^\bot}(\vec{x},t)$ changes the Lagrangian by the quantity $\delta L_p[\partial_t \vec{A}^\bot \rightarrow \partial_t \vec{A}^\bot + \delta \partial_t \vec{A}^\bot] = \int d\vec{x} \vec{\pi}(\vec{x},t) . \delta \partial_t \vec{A^\bot_p}(\vec{x},t)$. But since we change only the transverse part of the vector potential, we must have $\vec{\nabla} . \delta \partial_t \vec{A^\bot_p}(\vec{x},t) = 0$. So we can add to $\vec{\pi}(\vec{x},t)$ the gradient of a scalar function $f(\vec{x},t)$ without changing the variations $\delta L_p$. Indeed, 

\begin{eqnarray}
\delta L_p[\partial_t \vec{A}^\bot \rightarrow \partial_t \vec{A}^\bot + \delta \partial_t \vec{A}^\bot] &=& \int d\vec{x} (\vec{\pi}(\vec{x},t) + \vec{\nabla} f(\vec{x},t)) . \delta \partial_t \vec{A^\bot_p}(\vec{x},t) \nonumber \\
&=& \int d\vec{x} [\vec{\pi}(\vec{x},t). \delta \partial_t \vec{A^\bot_p}(\vec{x},t) -  f(\vec{x},t))  \vec{\nabla}. \delta \partial_t \vec{A^\bot_p}(\vec{x},t)] \nonumber \\
&=&  \int d\vec{x} \vec{\pi}(\vec{x},t). \delta \partial_t \vec{A^\bot_p}(\vec{x},t) \nonumber
\end{eqnarray}

The most general solution is $ \vec{\pi}(\vec{x},t) = -\varepsilon_0 \vec{E}(\vec{x},t) + \vec{\psi}(\vec{x},t)$ with $\vec{\psi}(\vec{x},t) =  \vec{\nabla} f(\vec{x},t)$. As shown in the following, the vector field $\vec{\psi}(\vec{x},t)$ is not a dynamical variable since it does not modify the equations of motion. The vector field $\vec{\psi}(\vec{x},t)$ contributes to a shift of the total energy. Fixing the reference of the energy to zero when all fields are null leads to the condition $\vec{\psi}(\vec{x},t) = \vec{0}$. Then $\vec{\pi}(\vec{x},t) = -\varepsilon_0 \vec{E}(\vec{x},t)$ as it is found in our paper\cite{rousseau} or in many books\cite{Weinberg,WeinbergField,Dirac,henneaux}.

\subsection{Calculation of the canonical momentum conjugated to $\vec{A}$ following Dirac procedure for constrained Hamiltonian \label{impl}}

As explained above, the Dirac procedure for constrained hamiltonian considers dynamical variables as being independent from each other\cite{Dirac}\cite[p.29]{henneaux}\cite[p.347]{WeinbergField}. Relationships are taken into account by a set a constraints denoted $\chi_i$ with $i=1,...,4$ in our paper\cite{rousseau}. For completeness, they are recalled in the table (\ref{table:comp}). This table shows that except for the constraint $\chi_2$ forgotten by G. K\'onya \textit{al.} \cite{2018arXiv180105590K} both results include the same list of constraints. Particularly, in our result also, the scalar and the vector potential are respectively given by the eq.(\ref{eq:Ap}) and the eq.(\ref{eq:phip}). Consequently, the independent degrees of freedom are exactly the same in both papers. As demonstrated above, the difference lies in the constraint $\chi_2$ which has not been taken into account in G. K\'onya \textit{al.} \cite{2018arXiv180105590K} but not from the consideration of different dynamical variables. 

Using the Dirac formalism for constrained hamiltonian, we show in the following that the proposition $\vec{\pi}' = -\varepsilon_0 \vec{E} -q\vec{r}\int_0^1 du \delta(\vec{x}-u\vec{r})$ as a momentum is excluded. From the set of equations (\ref{eq:Lpoin1}-\ref{eq:Lpoin3}) the canonical momentum conjugated to the vector potential is given by :

\begin{eqnarray}
\pi_i(\vec{x},t) = \frac{\delta L_p(\vec{r}, \vec{A}_p,\phi_p)}{\delta \partial_t {A}^i_p} = \varepsilon_0 [\partial_t A_i(\vec{x},t)+ \partial_i \phi(\vec{x},t)] = -\varepsilon_0 E_i(\vec{x},t)
\label{eq:pi}
\end{eqnarray}

As noted by G. K\'onya \textit{al.} \cite{2018arXiv180105590K} there is an ambiguity in the definition of the canonical momemtum $\vec{\pi}(\vec{x},t)$ see also ref.\cite[p.348]{WeinbergField} for more details. 

Changing $\partial_t \vec{A}(\vec{x},t)$ by the amount $\delta \partial_t \vec{A}(\vec{x},t)$ changes the Lagrangian by the quantity $\delta L = \int d\vec{x} \vec{\pi}(\vec{x},t) . \delta \partial_t \vec{A}(\vec{x},t)$. But since variations must also satisfy the gauge constraints, for example in the Coulomb gauge, we can add the gradient of a scalar function $f(\vec{x},t)$ without changing the variations $\delta L_c$. Indeed, 
$$\delta L_c = \int d\vec{x} [\vec{\pi}(\vec{x},t) + \vec{\nabla} f(\vec{x},t)] . \delta \partial_t \vec{A}_c(\vec{x},t)=   \int d\vec{x} \vec{\pi}(\vec{x},t).  \delta \partial_t \vec{A}_c(\vec{x},t) $$ 

In a similar fashion, in the Poincar\'e gauge, the vector-potential satifies $\vec{x}.\delta \vec{A}(\vec{x},t)  = x\vec{e}_r. \delta \vec{A}(\vec{x},t)= \delta A_r(\vec{x},t) = 0 $ where $A_r(\vec{x},t)$ is the component along the basis vector $\vec{e}_r$. So we can add any radial vector-field $\vec{\psi}(\vec{r})  = f(\vec{r}) \vec{e}_r$ without changing the Lagrangian variations $\delta L_p$:

$$\delta L_p = \int d\vec{x} [\vec{\pi}(\vec{x},t) + f(\vec{r}) \vec{e}_r ] . \delta  \partial_t \vec{A}_p(\vec{x},t)= \int d\vec{x} \vec{\pi}(\vec{x},t) . \delta  \partial_t \vec{A}_p(\vec{x},t)$$

Since the previous conditions have to hold at any time, we must have $\partial_t \vec{\psi}(\vec{r},t) = 0$ in both case. The field $\vec{\psi}(\vec{r})$ depends only on space variables.

At this stage, neither in the Coulomb gauge nor in the Poincar\'e gauge, the canonical momentum $\pi$ is uniquely defined by the eq.(\ref{eq:pi}). In the Poincar\'e gauge, in a generic way, it reads $\vec{\pi}(\vec{x},t) = \varepsilon_0 [\partial_t \vec{A}_p(\vec{x},t)+ \vec{\nabla} \phi_p(\vec{x},t) ] + \vec{\psi}(\vec{r}) $ where $\vec{\psi}(\vec{r})$ is a radial vector-field.

We can specify the vector-field $\vec{\psi}(\vec{x},t)$ with the help of the constraints and the Maxwell-Gauss equation. First, we need to find the Hamiltonian:

\begin{eqnarray}
H_p(\vec{r}, \vec{A}_p,\phi_p) &=& \vec{\mathcal{P}} . \dot{\vec{r}}  + \int d\vec{x} \vec{\pi}(\vec{x},t).\partial_t \vec{A}_p(\vec{x},t) - L_p(\vec{r}, \vec{A}_p,\phi_p) \nonumber \\
&=& \frac{1}{2m} [\vec{\mathcal{P}} - q \vec{A}_p(\vec{x},t)]^2 +V(\vec{r}) + q \phi_p(\vec{r},t) \nonumber \\
&+& \int d\vec{x} \{ \frac{1}{2\varepsilon_0}[ 2 \vec{\pi}^2(\vec{x},t) - (\vec{\pi}(\vec{x},t)-\vec{\psi}(\vec{x},t))^2 ] + \frac{1}{2\mu_0}\vec{B}^2(\vec{x},t) - \vec{\pi}(\vec{x},t).[\vec{\nabla} \phi_p(\vec{x},t) +\frac{ \vec{\psi}(\vec{x},t)}{\varepsilon_0}]\} \label{eq:Hpsi} 
\end{eqnarray}
where $ \vec{\mathcal{P}}$ is the canonical momentum associated to the particle position $\vec{r}$.

To obtain this expression one writes $\partial_t \vec{A}_p(\vec{x},t) = [\frac{1}{\varepsilon_0} \vec{\pi}(\vec{x},t) - \vec{\nabla} \phi_p(\vec{x},t) - \frac{1}{\varepsilon_0} \vec{\psi}(\vec{x})]$

At this step, all dynamical variables $\vec{A}_p(\vec{x},t), \vec{\pi}(\vec{x},t)$, $\phi(\vec{x},t)$, $\pi_\phi(\vec{x},t)$ are assumed to be independent. Since $\frac{\delta L}{\delta \partial_t \phi} = \pi_\phi = 0$, we have found one constraint that applies to the dynamics. This constraint has to hold at any time. So, the following should hold:

\begin{eqnarray}
\partial_t \vec{\pi}_\phi(\vec{x},t) = \{ \pi_\phi, H_p\} = -\frac{\delta \pi_\phi}{\delta \pi_\phi}\frac{\delta H_p}{\delta \phi} = -[q \delta(\vec{x}-\vec{r}) + \vec{\nabla}.\vec{\pi}(\vec{x},t)] = 0
\label{eq:cons}
\end{eqnarray}

The Hamiltonian $H_p$ is given by the equation (\ref{eq:Hpsi}).

On the other hand, the Maxwell-Gauss equation has to hold too: $q \delta(\vec{x}-\vec{r}) - \varepsilon_0 \vec{\nabla}.\vec{E}(\vec{x},t) = 0$ leading to the constraint $\vec{\nabla}.\vec{\psi}(\vec{x}) = 0$ for the field $\vec{\psi}(\vec{x})$.

To conclude, there is a freedom in the choice of the canonical momentum associated to the vector potential but with some constraints as summarized in table (\ref{table:psi}). \\

\begin{table}[ht]
\centering
\begin{tabular}{p{6cm} p{6cm}}
\begin{eqnarray}
&& \text{Poincar\'e~ gauge}  \nonumber \\
&&\vec{\pi}(\vec{x},t) = -\varepsilon_0 \vec{E}_p(\vec{x},t) + \vec{\psi}(\vec{x}) \nonumber \\
&&\text{with~}  \vec{\psi}(\vec{x}) = f(\vec{x}) \vec{e}_r ~,~\partial_t \vec{\psi}(\vec{x}) = 0 \nonumber \\ 
&&\text{~and~} \vec{\nabla}.\vec{\psi}(\vec{x}) = 0 \nonumber 
\end{eqnarray}
&
\begin{eqnarray}
&& \text{Coulomb~gauge} \nonumber \\
&& \vec{\pi}(\vec{x},t) = -\varepsilon_0 \vec{E}_p(\vec{x},t) + \vec{\psi}(\vec{x}) \nonumber \\
&&\text{with~}  \vec{\psi}(\vec{r}) = \vec{\nabla}f(\vec{x})  ~,~\partial_t \vec{\psi}(\vec{x}) = 0 \nonumber \\ 
&&\text{~and~} \vec{\nabla}.\vec{\psi}(\vec{x}) = 0 \nonumber 
\end{eqnarray} 
\end{tabular}
\caption{Conditions satisfied by the vector field $\vec{\psi}$ in the Poincar\'e and in the Coulomb gauge.}
\label{table:psi}
\end{table}

The condition $\vec{\nabla}.\vec{\psi}(\vec{x}) = 0$ does not depend on the chosen gauge. It results from $\partial_t \pi_\phi =0$.
As a consequence there is here an additional argument against G. K\'onya \textit{al.} \cite{2018arXiv180105590K} proposition [eq:(\ref{eq:piK})] as a canonical momentum $\vec{\pi}$. The vector field $\vec{\psi}(\vec{x}) = -q\vec{r}\int_0^1 du \delta(\vec{x}-u\vec{r}) $ that they proposed is not divergence-free. So the constraint $\chi_2$ given by the eq.(\ref{eq:cons}) excludes this solution. Again, the constraint $\partial_t \pi_\phi =0$ excludes G. K\'onya \textit{al.} \cite{2018arXiv180105590K} result as a valid solution for the canonical momentum  $\vec{\pi}$.

For completeness we specify a bit more the radial field $\vec{\psi}$ and show that it does only change the reference of the energy. For this we need the Hamiltonian. The Dirac procedure for constrained hamiltonian can be continued as in our paper\cite{rousseau}. The Dirac brackets can be computed\cite{rousseau}. After these computations the constraints act effectively. The hamiltonian can be simplify by taking all the constraints into account. After, integration by part, it reads:

\begin{eqnarray}
H_p(\vec{r}, \vec{A}_p,\phi_p) &=& \frac{1}{2m} [\vec{\mathcal{P}} - q \vec{A}_p(\vec{x},t)]^2 + V(\vec{r})  \nonumber \\
&+& \int d\vec{x} \{ \frac{\vec{\pi}^2(\vec{x},t)}{2\varepsilon_0} + \frac{\vec{\pi}(\vec{x},t).\vec{\psi}(\vec{x},t)}{\varepsilon_0} -\frac{\vec{\psi}^2(\vec{x},t)}{2\varepsilon_0} ] - \frac{\vec{\pi}(\vec{x},t). \vec{\psi}(\vec{x},t)}{\varepsilon_0} + \frac{1}{2\mu_0}\vec{B}^2(\vec{x},t) ]\} \nonumber \\
&+& \int d\vec{x}  \phi_p(\vec{x},t)[\vec{\nabla}.\vec{\pi}(\vec{x},t) + q\delta(\vec{x}-\vec{r})] \nonumber
\end{eqnarray}

It can be simplified with the help of the constraint $\chi_2$ [eq:(\ref{eq:cons})].

\begin{eqnarray}
H_p(\vec{r}, \vec{A}_p,\phi_p) &=& \frac{1}{2m} [\vec{\mathcal{P}}  - q \vec{A}_p(\vec{x},t)]^2 + V(\vec{r})  \nonumber \\
&+& \int d\vec{x} \{ \frac{\vec{\pi}^2(\vec{x},t)}{2\varepsilon_0}   + \frac{1}{2\mu_0}\vec{B}^2(\vec{x},t) \} \nonumber \\
&-& \int d\vec{x} \frac{\vec{\psi}^2(\vec{x},t)}{2\varepsilon_0} 
\end{eqnarray}

This last expression shows that the field $\vec{\psi}(\vec{x},t)$ is not a dynamical variable. It does not contribute to any equations of motion. Actually, it just adds up a constant contribution to the total energy. But the energy can only be defined up to a constant. If we make the usual choice $H=0$ as the origin of the energy when there is no electromagnetic field then $\vec{\psi}(\vec{x},t)=0$. we recover then the usual results:

\begin{eqnarray}
H_p(\vec{r}, \vec{A}_p,\phi_p) &=& \frac{1}{2m} [\vec{\mathcal{P}}  - q \vec{A}_p(\vec{x},t)]^2 + V(\vec{r})  + \int d\vec{x} \{ \frac{\vec{\pi}^2(\vec{x},t)}{2\varepsilon_0}   + \frac{1}{2\mu_0}\vec{B}^2(\vec{x},t) \}  
\label{eq:Hpoin} \\
\vec{\pi}(\vec{x},t) &=& -\varepsilon_0 [\partial_t \vec{A}_p(\vec{x},t)+ \vec{\nabla} \phi_p(\vec{x},t) ] =  -\varepsilon_0  \vec{E}(\vec{x},t)
\end{eqnarray}

As a conclusion, the Power-Zienau-Woolley hamiltonian is not the minimal-coupling hamiltonian written in the Poincar\'e gauge. As previously shown\cite{Kobe1988}, it remains form-invariant through a gauge transformation.
 
\section{Inserting $A_p^\perp = A_C$ and $A_p^\parallel$ in our result does not lead to the Power-Zienau-Woolley Hamiltonian}

On the contrary to K\'onya \textit{et al.}\cite{2018arXiv180105590K} we do not consider that our choice of dynamical variables is mandatory. It just considers the vector potential as a whole quantity. Nevertheless as in K\'onya \textit{et al.}\cite{2018arXiv180105590K} paper the transverse part of the vector potential is the only independent variable. Nonetheless once our result is known one is always free to exhibit the only independent variable and can write $\vec{A}_p = \vec{A}_p^\bot + \vec{A}_p^\parallel[ \vec{A}_p^\bot]$. So we do not share the conclusion that our choice of variable is an "awkward choice". It has been described as an "awkward choice" by K\'onya \textit{et al.} based on a quote from Weinberg's book\cite{Weinberg}. In fact, K\'onya \textit{et al.} quotation of Weinberg writings is approximative and changes its very meaning. Page 15 of their manuscript\cite{2018arXiv180105590K}, they wrote: \textit{"As explained by Weinberg in Section 11.3 of his book [13], $\Pi'_C=-\varepsilon_0 E$ is "an awkward choice" (quote: Weinberg) for the canonical field momenta because when quantized, it does not commute with the particle momenta"}

Here is the exact citation\cite[p.315-316]{Weinberg}:
\textcolor{blue}{There is an awkward feature about the canonical commutation relations in Coulomb gauge, that we have not yet uncover. Although the commutators of the particle coordinates $x_{n j}$ with $A_i$ and  $\Pi_i$ all vanish, the particle momenta $p_{n j}$ have non-vanishing commutators with $\Pi_i$. According to the Dirac prescription and Eqs. (11.3.8)-(11.3.11), this commutator is [not zero]. We can avoid this complication by introducing as a replacement for $\Pi_i$  its solenoidal part}.
 
The "awkward feature" is actually not related to the choice of the dynamical variables but to a transition to the interaction picture. Weinberg made another statement in Ref.\cite[p.348-349]{WeinbergField} confirming our understanding of his previous citation:
\textcolor{blue}{Although the commutation relations (8.3.5) [commutation relations with A and $\Pi$ involving the Dirac-transverse distribution] are reasonably simple, we must face the complication that $\Pi$ does not commute with matter fields and the canonical conjugates. If F is any functional of these matter degrees of freedom, then its Dirac bracket with A vanishes, but its Dirac brackets with $\Pi$ is [not zero]...In order to facilitate the transition to the interaction picture, instead of expressing the Hamiltonian in terms of A and $\Pi$, we shall write it in terms of A and $\Pi_\perp$, where $\Pi_\perp$ is the solenoidal part of $\Pi$.} 

Weinberg then expresses the dynamical variables in terms of the longitudinal and transverse part. He then writes $\vec{A} = \vec{A}^\perp + \vec{A}^\parallel$ and $\vec{\pi} = \vec{\pi}^\perp + \vec{\pi}^\parallel$ and inserts these expressions into the commutators and the hamiltonian previously derived. 

So can the Power-Zienau-Woolley hamiltonian be obtained by writing $\vec{A}_p = \vec{A}_p^\perp + \vec{A}_p^\parallel$ and $\vec{\pi}_p = \vec{\pi}_p^\perp + \vec{\pi}_p^\parallel$ in our Hamiltonian [eq.(8) of the main manuscript or equation (\ref{eq:Hpoin}) in these paper]?

Of course not. If we do the same procedure as Weinberg's and write $\vec{A}_p = \vec{A}_p^\perp + \vec{A}_p^\parallel$ and $\vec{\pi}_p = \vec{\pi}_p^\perp + \vec{\pi}_p^\parallel$, we do not recover the Power-Zienau-Woolley hamiltonian since  $\vec{\pi}_p^\perp = -\varepsilon_0 \vec{E}^\perp  \neq \vec{D}$ where $\vec{D}$ is the displacement vector as shown previously. 

As a conclusion, the Power-Zienau-Woolley hamiltonian \underline{cannot} be derived from the minimal-coupling hamiltonian through a gauge transformation. 

\section{Some weaknesses of the Power-Zienau-Woolley hamiltonian}

The Power-Zienau-Woolley hamiltonian reads:

\begin{eqnarray}
H_{PZW} &=& \frac{1}{2m}[\mathcal{P}+q \vec{r} \times \int_0^1 du u \vec{B}(u\vec{r},t)]^2 \\
&+& \int d^3 x \frac{1}{2 \varepsilon_0} \vec{D}^2(\vec{x},t) + \frac{1}{2 \mu_0} \vec{B}^2(\vec{x},t) \\
&-& - \frac{1}{\varepsilon_0} \int d^3 x  ~\vec{D}(\vec{x},t).\vec{P}(\vec{x},t) \\
&+& + \frac{1}{2 \varepsilon_0} \int d^3 x ~ \vec{P}^2(\vec{x},t)
\end{eqnarray}

This is the equation eq:(46) in K\'onya \textit{et al.} comment\cite[p.14]{2018arXiv180105590K} and according to them this is also the minimal coupling hamiltonian in the Poincar\'e gauge. They made the following comments quote in italic:

\begin{enumerate}
\item \textit{The PZW Hamiltonian is free from the A-square term},

Maybe it is here a question of semantic. But it can not be said that the PZW hamiltonian is free from the A-square term since precisely $\vec{r} \times \int_0^1 du u \vec{B}(u\vec{r},t) = -\vec{A}_p(\vec{r},t)$ is the vector potential in the Poincar\'e gauge. This term is usually neglected in the so-called electric-dipole approximation but it does contribute in the complete theory.

\item \textit{accounts for the light-matter interaction in the form of the $\vec{D}(\vec{x},t).\vec{P}(\vec{x},t)$ term},
\item \textit{contains a P-square term.}

By definition $\vec{D}(\vec{x},t) = \varepsilon_0\vec{E}(\vec{x},t) + \vec{P}(\vec{x},t)$. Replacing this definition into the set of equations (1-4), one can remark that the so-called light-matter interaction term and the P-square term cancel out. It is dramatic in the electric-dipole approximation since there is then no interaction term. Indeed, the Power-Zienau-Woolley Hamiltonian reduces to:

\begin{eqnarray}
H_{PZW} & \simeq & \frac{1}{2m}\mathcal{P}^2 + \int d^3 x ~ \frac{1}{2 \varepsilon_0} \vec{E}^2(\vec{x},t) + \frac{1}{2 \mu_0} \vec{B}^2(\vec{x},t)
\nonumber 
\end{eqnarray}

\end{enumerate}

Those criticisms previously raised in our paper weaken strongly the validity of the Power-Zienau-Woolley Hamiltonian. Nevertheless K\'onya \textit{et al.} did not comment on them.

\section*{Conclusion}

To conclude, we have shown that if all the constraints are taken into account correctly then the canonical momentum $\vec{\pi}(\vec{x},t)$ conjugated to the vector potential is $-\varepsilon_0 \vec{E}(\vec{x},t)$ provided that the reference of the energy is taken to be null. This result has been derived following K\'onya \textit{et al.} methodology\cite[p.14]{2018arXiv180105590K} where all quantities are written with the help of the independent dynamical variable $\vec{A}^\bot_p(\vec{x},t)$. We have shown that the constraint $\partial_t \pi_\phi = 0$ leads to this result.  We have also recalled our derivation based on the Dirac theory for constrained hamiltonian. We have obtained the same result  $\vec{\pi}(\vec{x},t) = -\varepsilon_0 \vec{E}(\vec{x},t)$ based on the same argument $\partial_t \pi_\phi = 0$. Moreover, this derivation allowed us to conclude that the following proposition for the momentum $\vec{\pi}'(\vec{x},t) = -\varepsilon_0 \vec{E}(\vec{x},t) -q\vec{r}\int_0^1 du \delta(\vec{x}-u\vec{r})$  cannot be considered as correct since $-q\vec{r}\int_0^1 du \delta(\vec{x}-u\vec{r})$ is not divergence-free. We have also explained that the differences between our both results cannot be attributed to the consideration of different dynamical variables since they are similar in both papers. The independent dynamical variables are taken into account explicitly in K\'onya \textit{et al.} \cite{2018arXiv180105590K} work and implicitly in our work through the constraints $\chi_i$. Nevertheless they are exactly the same.

In order to obtain the Power-Zienau-Woolley hamiltonian one needs $\vec{\pi}'(\vec{x},t) = - \vec{D}(\vec{x},t)$, which is impossible as we have demonstrated. As a consequence, we conclude that this hamiltonian cannot be derived from the minimal-coupling hamiltonian. We ended these notes in highlighting some weaknesses of the Power-Zienau-Woolley hamiltonian. More weaknesses can be found in our paper. Particularly the physical meaning of the term $\frac{1}{2 \varepsilon_0} \vec{D}^2(\vec{x},t) + \frac{1}{2 \mu_0} \vec{B}^2(\vec{x},t)$ is questionable since this is neither the electromagnetic-field energy-density in vacuum nor in matter. As far as we know, the weaknesses of the Power-Zienau-Woolley hamiltonian have never been commented (and answered) into the literature.

\bibliographystyle{nature-doi}
\bibliography{biblio_Quantum_Hamiltonian_Reply}

\section*{Author contributions statement}

DF initiated this work. ER did the calculations. ER and DF have discussed the results and reviewed the manuscript. 

\section*{Additional information}

The authors declare no competing financial interests.

The corresponding author is responsible for submitting a \href{http://www.nature.com/srep/policies/index.html#competing}{competing financial interests statement} on behalf of all authors of the paper. This statement must be included in the submitted article file.

\end{document}